\documentclass[prb,twocolumn,aps,showpacs]{revtex4}
\usepackage{natbib}
\usepackage{graphicx}
\usepackage{amsfonts}
\usepackage{bm}

\begin{document}
\title{Nonlinear conductivity of two-dimensional Coulomb glasses}
\author{M. Caravaca, A.\ M.\ Somoza and M.\ Ortu\~no}
\affiliation{Departamento de F\'{\i}sica - CIOyN, Universidad de Murcia, 
Murcia 30.071, Spain}

\begin{abstract}
We have studied the nonlinear conductivity of two-dimensional Coulomb glasses. 
We have used a Monte Carlo algorithm 
to simulate the dynamic of the system under an applied electric field $E$.
We found that in the nonlinear regime the site occupancy in the Coulomb gap
follows a Fermi-Dirac distribution with an effective temperature $T_{\rm eff}$,  
higher than the phonon bath temperature $T$. 
The value of the effective temperature is compatible with that obtained for slow
modes from the generalized fluctuation-dissipation theorem.
The nonlinear conductivity for a given electric field and $T$ is fairly similar to the 
linear conductivity at the corresponding $T_{\rm eff}$.
We found that the dissipated power and the effective temperature are related by 
an expression of the form  $(T_{\rm eff}^\alpha-T^\alpha)T_{\rm eff}^{\beta-\alpha}$.
\end{abstract}
\maketitle

\section{Introduction}

Electron transport in Coulomb glasses (CG) has been investigated for decades.
These glasses are systems with electronic states localized by the disorder 
and long range Coulomb interactions between carriers. 
At low temperatures, conductivity in CG is by  hopping, where 
the transition rates for electron jumps depend
exponentially on an energy factor and on a spatial factor. In the limit of very
low temperatures, an optimization of the total penalty paid through these
two factors leads to the mechanism 
coined by Mott as variable range hopping (VRH). Mott obtained
the precise law for the DC conductivity $\sigma$ of non--interacting
systems in this regime \cite{Mo68} and
Efros and Shklovskii (ES) \cite{EfSh75} modified his argument to include the
effects of Coulomb interactions by considering the specific form of the
single--particle density of states in CG.
The conductivity in this case is of the form
\begin{equation}
\sigma \propto \exp \lbrace - \left( T_0/T\right)^{1/2} \rbrace
\label{unmedio}
\end{equation}
with the exponent  $1/2$ independent of the dimensionality of the
system. $T_0=\beta {e^2}/({\epsilon k\xi})$ is a characteristic temperature,
$\epsilon$ the dielectric constant of the material, $k$ Boltzmann constant, $\xi$ the localization 
radius of the electrons, and $\beta$ a
numerical coefficient that depends on dimensionality.
From percolation theory for hopping transport one obtains
$\beta \approx 6.5$ for 2D systems (see references in \onlinecite{LeNg87}).
The conductivity of many different types of systems have been found to obey
this law, Eq.\ (\ref{unmedio}). 

Nonlinear effects in electron transport are specially important in CG, where
interactions usually increase nonlinearities and can also establish an
effective temperature for the electrons, higher than the phonon bath temperature,
at very low temperatures when the thermal conductance between the electrons and 
the phonons is not large enough to dissipate all the electrical power provided to
the system.
While there are many experimental studies of nonlinear effects on systems showing
VRH in the presence of  a Coulomb gap, there is no proper theory addressing the problem.
Some experimental results have been interpreted in terms of theoretical approaches
designed for the noninteracting VRH regime, and extended to the interacting case
by adapting the typical hopping length.
For electric fields satisfying the condition $E_{\rm c}<E<kT/el_0$, where 
$E_{\rm c}=kT/eL$, $l_0$ is the average
distance between nearest neighbor impurities and $L$ is some typical hopping length,
the theoretical models \cite{PoRi76,Sh76} predict a conductivity of the form
\begin{equation}
\sigma(T,E)=\sigma(T,0)\exp\left\{eEL/kT\right \}.\label{field}
\end{equation}
The different models differ in the precise form of the typical hopping length $L$.
While the argument of Pollak and Riess \cite{PoRi76} adapted to VRH in CG would produce
$L\propto T^{-1/2}$, that of Shklovskii \cite{Sh76} would result in $L\propto T^{-1}$.
These approaches were called field-effect models and where relatively successful in explaining early experimental results.  
Some recent experiments can also be interpreted in terms of these models, for example, the results of Grannan {\it et al.} \cite{GrLa92}, which fit well to  $L\propto T^{-1}$,
and those of Zhang {\it et al.} \cite{ZhCu98a}, which fit adequately to $L\propto T^{-1/2}$.

In 1990, Wang {\it et al.} \cite{WaWe90} presented thorough results on neutron-transmutation 
doped Ge that could not be fitted with any field-effect model and which could be explained with the so called hot-electron model. This model assumes that the applied electric power is deposited in the electron system and is dissipated to the phonon system through the electron-phonon coupling.
For slowly relaxing systems in general, and hopping systems at very low temperatures in particular, this coupling is too weak to be able to thermalize together the electron and the phonon systems. 
The electron-electron interaction establishes an effective temperature $T_{\rm eff}$ for the electron system, which remains higher than the phonon bath temperature $T$. 
The model also assumes that the conductivity for any field $E$ and temperature $T$ only depends
on the electron system and is the same  as the linear conductivity at the effective temperature    
\begin{equation}
\sigma(T,E)=\sigma(T_{\rm eff},0) \propto \exp \lbrace - \left( T_0/T_{\rm eff}\right)^{1/2} \rbrace .
\label{unmedioeff}
\end{equation}
The effective temperature can be calculated from the thermal conductivity between the conducting electrons and the phonon bath. This quantity is not known for hopping systems and by analogy with metals one assumes that it is proportional to a power of the temperature. Then the effective temperature is related to the electric power $P$ through
\begin{equation}
P=a(T_{\rm eff}^\beta-T^\beta) 
\label{power}
\end{equation}
where $a$ is independent of temperature.

Zhang {\it et al.} \cite{ZhCu98a} performed a systematic study of non linear effects on 
doped Si and Ge and concluded that field-effect models explain adequately the experimental 
results when $T_0/T>135$, while hot-electron models provide a better fit when $T_0/T<135$. 
In line with these results, Gershenson {\it et al.} \cite{GeKh00} studied two-dimensional 
hopping and concluded that in systems with small localization lengths (and so large $T_0$)
field-effects dominate, while in systems with large localization lengths the hot-electron model
explains adequately the results.
Recently, many experimental results have been fitted with the hot-electron model
\cite{LeLh03,MiSh04,GaLi07,JaRa08,FiGe09}, while some still have been interpreted in terms of field-effect models \cite{LaLh00,YuWa04}.

The hot-electron model together with a strong temperature dependence of the conductivity
presents an instability at very low temperatures. As the applied voltage is increased the
current may abruptly change by several orders of magnitude, as it has been observed in amorphous indium oxide films in the insulating state \cite{OvSa09}. This effect has been explained in terms of a bistability of the effective temperature \cite{AlKr09}.  We will point out later on that a similar effect could also be present in Coulomb glasses at sufficiently low temperatures \cite{LaSa96}. 

In this paper, we perform Monte Carlo (MC) simulations of CG in the nonlinear
regime and calculate the conductivity as a function of temperature and applied field.
As we can measure the effective temperature of the electrons directly \cite{SoOr08} and the absorbed and emitted powers separately, our results are an excellent tool to check the validity 
of the hot-electron model. 
After presenting our model in the next section, we study the effective temperature $T_{\rm eff}$ as obtained from the site occupation and analyze its relation with the temperature of slow modes obtained from the generalization of the fluctuation-dissipation theorem. In section IV, we present the results for the nonlinear conductivity in the variable range hopping regime as a function of the temperature and the electric field and analyze the applicability of the hot-electron model. In section V, we study the dissipated power in the non linear regime in terms of  $T_{\rm eff}$. In section VI, we analyze the predictions of the hot-electron model. We finalize with some conclusions.

\section{Model}

To calculate the conductivity of two--dimensional  Coulomb
glasses we consider the Hamiltonian
\cite{PoOr85}
\begin{equation}
H=\sum_i(\phi_i+x_{i}E) n_i +\sum_{i<j} \frac{(n_i-K)(n_j-K)}{r_{ij}}\;, \label{hamil}
\end{equation}
where $n_i=0,1$ are occupation numbers, $K$ is the compensation, equal to 1/2, and $\phi_i$ 
are the random site energies chosen
from a box distribution with interval $[-W/2, W/2]$. $E$ is the value of the applied electric field. $x_i$ is the coordinate along the direction of the applied field of site
$i$ and $r_{ij}$ is the distance between sites $i$ and $j$.
We consider  square samples of lateral size $L$ and with $N$ sites
placed at random with a minimum separation between them of $0.2 l_0$. 
$l_0=L/\sqrt{N}$ is our unit of distance and $1/l_0$ is our
unit of energy and temperature.
We study systems ranging from 500 to 4000 sites with a range of disorder $W=2$
and two values of the localization length, $\xi=1$ and 2.
We implement periodic boundary conditions and Coulomb interactions are calculated 
using the minimum image convention \cite{DaLe84},
which is a reasonable choice for disordered systems in order to reduce finite size effects
 avoiding at the same time artificial long-range correlations.
The electric field energy increases linearly in the direction of the applied field. 
Once periodic boundary conditions are imposed, the total energy of the system is not well-defined, but the transition energies are well-defined provided that we restrict ourselves to hops of length shorter than half the lateral size of the system.
The previous arrangement induces 
a permanent current in the direction of the field when we set up our dynamical procedure.

We concentrate here in the regime where the temperature 
is low enough for conduction to be by variable range hopping, and high enough so that
a stationary state is achieved relatively fast.
In this case we can consider only single-electron transitions, since we are not
in the regime where  many-electron transitions are important  \cite{SoOr06}.
The clusterization algorithm previously developed by us \cite{SoOr05} is not
efficient in the temperature range considered here.  
As we are studying stationary states, not equilibrium properties, a kinetic MC algorithm
that keep track of the physical time taken in the different processes must be used
\cite{KwLa04}.
The MC method employed first chooses a pair of sites with a probability
proportional to $\exp ({-2 r_{ij}}/{\xi})$ \cite{TsPa03}. 
Then if one of the sites chosen is occupied and the other empty it exchanges their occupation
 when the total transition energy $\Delta E$, including the
energy due to the applied field, is negative or with probability $\exp ({-\Delta E}/T)$
when $\Delta E >0$. 
The time step of our MC procedure is then equal to
$\tau_0/\sum_{ij} \exp ({-2 r_{ij}}/{\xi})$, where $\tau_0$ is the inverse phonon 
frequency, of the order of $10^{-13}$ s \cite{BoKa75,MoTh97}.

In the non linear regime we define the conductance as the ratio of the intensity
divided by the applied voltage, $V=EL$. As we consider square samples, the conductivity is thus $\sigma=I/(EL)$.
We start from a random configuration and follow the dynamics 
at a given temperature, monitoring all relevant magnitudes.
Once we are in a stationary situation, we obtain the conductivity of each sample
through the displacement of the center of mass of the electrons for a given time interval \cite{CaVo09}.
The temperature range studied goes from 0.05 to 0.2, the typical number of samples is
5000 and the simulation time is $10^7\tau_0$. 
The values of the applied electric field run up to $2T$.

\section{Effective temperature}

With our MC procedure we simulate the dynamics of CG in the presence of a relatively
strong electric field, of the order of $T$, so that we are in the nonlinear regime.
Once the system has reached a stationary state, we calculate the site occupation probability 
$f(\epsilon_i)$ as a function of site energy $\epsilon_i$, which is defined as
\begin{equation}
\epsilon_i=\phi_i +\sum_{j\ne i} \frac{(n_j-K)}{r_{ij}}\;. \label{senergy}
\end{equation}
We note that this energy does not include the electric field contribution.
We find that $f(\epsilon_i)$ follows pretty well
a Fermi-Dirac distribution with an effective temperature $T_{\rm eff}$, which depends
on both $T$ and the applied field $E$. 
In Fig.\ 1 we represent the occupation probability of sites near the chemical potential as a function of site energy for $T=0.1$ and two different values of the electric field $E=T/2$ (circles) and
$T$ (squares). The continuous lines are Fermi-Dirac distributions corresponding to 
$T_{\rm eff}=0.112$ and $0.136$, respectively.
\begin{figure}[htb]
\includegraphics*[width=.45\textwidth]{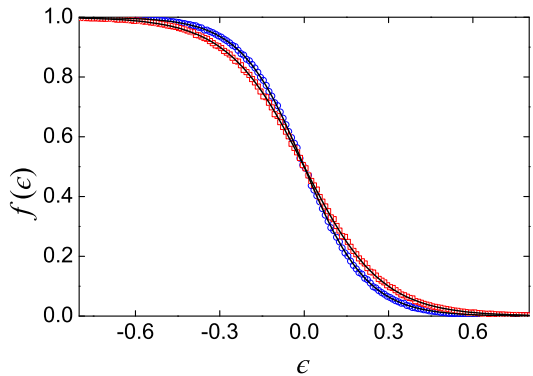}
\caption{%
(Color online) Occupation distribution function near the chemical potential  at $T=0.1$ for two values of the applied field $E=T/2$ (blue circles) and $E=T$ (red squares). The continuous lines are Fermi-Dirac distributions for temperatures 0.112 and 0.136,
respectively.} \label{fig1}
\end{figure}

Let us call $n(\Delta E)$ the density of electron-hole excitations of energy $\Delta E$ and longer than a given length, that we take equal to 10. 
The relative probability of having long electron-hole excitations with negative and positive energies, $n(-\Delta E)/(n(-\Delta E)+n(\Delta E))$, follows a distribution very similar to the site energy occupation probability. The situation is similar to simulations of relaxation \cite{SoOr08}.
Long jumps and injection or extraction of single particles near the Fermi Level are both very slow process equilibrated at the same effective temperature. 
In practice,
it is easier to calculate $T_{\rm eff}$ from the distribution of the excitations than from the site occupation function, since for the former the statistics are better
and the procedure is insensitive to the determination of the chemical potential \cite{SoOr08}. 

The quality of the fit of the data in Fig.\ 1 to a FD distribution indicates that the electron-electron
interaction has indeed thermalized the electrons near the chemical potential at the effective temperature. However, we note that not all electrons
are thermalized at $T_{\rm eff}$. Fast electronic modes, like short electron-hole excitations, are equilibrated at $T$, while slow modes, like long excitations, are equilibrated at $T_{\rm eff}$.

The existence of two different temperatures in the electronic system can be also observed
through the extension of the fluctuation-dissipation theorem for non-equilibrium systems in the way described in the spin glass literature \cite{CuKu97a,Ku05}.
In order to compare our results with those based on the fluctuation-dissipation theorem, we follow the same procedure as in Refs.\ \onlinecite{Gr04a,KoGr05}.
We study the response of the system to an external perturbation of the form $\delta\phi_i=\lambda(t)\varphi_i=\lambda_0\varphi_i\theta(t-t_w)$, where $\lambda_0\ll 1$ and $\varphi_i$ are normalized random variables uncorrelated from site to site and from the original random site energies.
The quantity conjugated to $\lambda(t)$ is $\delta n(t)\equiv 1/N\sum_i \langle \delta n_i(t)\varphi_i \rangle$, where $\delta n_i(t)=n_i(t)-K$. 
In the linear response regime the latter quantity is given by
\begin{equation}
\delta n(t)=\lambda \chi (t+t_w,t_w),
\end{equation}
where $\chi$ is the local susceptibility. At equilibrium, $\chi$ is related to the local charge correlation function
\begin{equation}
C(t+t_w,t_w)=\frac{4}{N}\sum_i\langle\delta n_i(t+t_w)\delta n_i(t_w)\rangle,
\end{equation}
through the fluctuation-dissipation theorem $T\chi(t)=1-C(t)$.
Out of equilibrium, this relation between $\chi$ and $C$ is not verified in general, but
in slowly relaxing systems it is often found that at short time scales the equilibrium relation still holds, while at long time scales there is also a linear relation with a different slope. From this slope we can define an effective temperature through $\partial \chi/\partial C = -1/T_{\rm eff}$ \cite{Ku05}.

\begin{figure}[ht]
\includegraphics*[width=.45\textwidth]{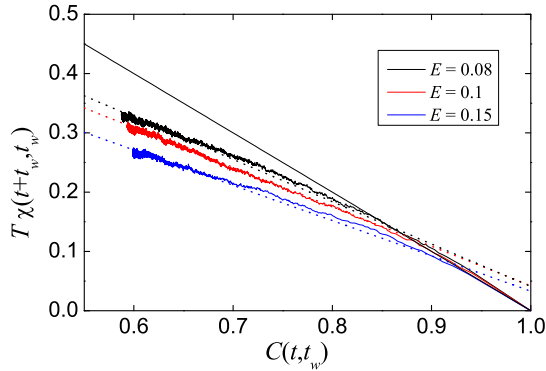}
\caption{%
(Color online) $T\chi (t+t_w,t_w)$
as a function of the correlation function for $E=0.008$ $0.1$ and $0.15$, from top to bottom,
and $T=0.1$. 
The dashed lines have slopes equal to 
$-T/T_{\rm eff}$ as obtained from the excitation probabilities.} \label{fig2}
\end{figure}

We calculate $C(t+t_w,t_w)$ and $\chi (t+t_w,t_w)$ for several values of the applied electric field
and in Fig.\ \ref{fig2} we present parametric plots of $T\chi (t+t_w,t_w)$
versus $C(t+t_w,t_w)$ for $T=0.1$. The system is out of equilibrium in a stationary state due to
an applied electric field, and so $t_w$ is irrelevant in this case. Each curve in Fig.\ \ref{fig2}  corresponds to a different electric field $E=0.008$, $0.1$ and $0.15$, from top to bottom. 
All curves start, for short time intervals, on the lower right corner with an slope equal one and then flatten out on the left part, for long time intervals,
 presenting a slope smaller than one and so an effective temperature higher than $T$.
The dashed lines have slopes equal to  $T/T_{\rm eff}$, where the effective temperature has been calculated from
the excitation probabilities as described above. 

In Fig.\ \ref{fig2} we can appreciate a fairly well defined effective temperature for slow processes, 
much better defined
than in the case of relaxation from high energy states \cite{SoOr08}. 
We also note the good agreement between the $T_{\rm eff}$
obtained from the long excitation probabilities and from the extension of the fluctuation-dissipation theorem.
We would like to add that the use of the latter approach is a noisy 
procedure that requires averages over many samples, much less convenient 
than our method, although this
needs the explicit identification of slow processes. In spin glasses, 
for example, we have not been able to apply our method so far.
Coulomb glasses are very suitable to study the role of an effective temperature since we have two types of simple slow processes: the
changes in occupation of sites near the chemical potential and long electron-hole excitations \cite{SoOr08}.

\begin{figure}[htb]
\includegraphics*[width=.45\textwidth]{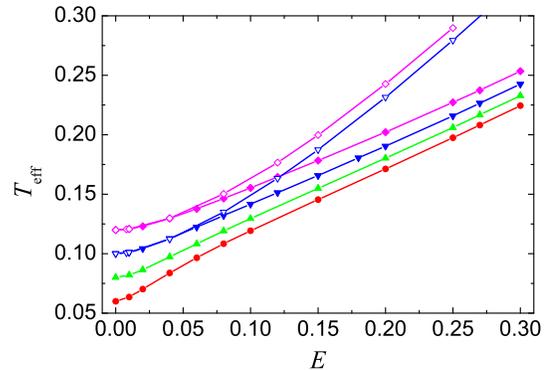}
\caption{%
(Color online) Effective temperature as a function of the applied electric field for several temperatures $T=0.06$ (red circles), 0.08 (green up triangles), 0.1 (blue down triangles) and 0.12 (magenta diamonds). 
Solid and empty symbols correspond to $\xi=1$ and $2$, respectively.} \label{fig3}
\end{figure}

We have systematically studied the effective temperature
in terms of the electric field and temperature.
From now on we will refer to the effective temperature obtained through the density
of excitations, which is the most precise and less noisy procedure.
In Fig.\ \ref{fig3} we plot  $T_{\rm eff}$ as a function of $E$ for several values of the temperature $T=0.06$ (circles), 0.08 (up triangles), 0.1 (down triangles) and 0.12 (diamonds).
We have also considered two values of the localization length $\xi$.
Solid symbols correspond to  $\xi=1$ and empty symbols to $\xi=2$. The lines are a guide to the eye.
These data cannot be obtained experimentally, thus our direct calculation of $T_{\rm eff}$ is a powerful tool to evaluate the validity of the hot-electron model.
After we study the non linear conductance, we will come back to this issue.

In the presence of an electric field, the electronic system can increase its energy by long hops along the direction of the field. Part of this energy is  very quickly dissipated via phonon emission by fast degrees of freedom, like soft dipoles and possible relaxation of
the electron that performed the long jump. Due to the Coulomb interaction, the energy provided
by the field is distributed over the whole electronic system and part of it
excites slow degrees of freedom. The system (more specifically, its slow degrees of freedom) cannot relax fast enough and remains hot. In the absence of slow degrees of freedom, one would expect an effective temperature $T_{\rm eff}=E\xi/2$ (for large enough values of the electric field) as derived in Ref.\ \onlinecite{MaSh92}.
The linear dependence between $T_{\rm eff}$ and $E$ at high fields in Fig.\ \ref{fig3} may correspond to this situation. The slopes of these straight segments for $\xi=1$ are indeed very close to 0.5, while the slopes for $\xi=2$ are 0.95 in agreement with this theoretical prediction. This model cannot explain the behavior for small fields, in particular the region for intermediate fields where $T_{\rm eff}$ is almost independent on $\xi$, and the change in curvature for low fields and low temperatures.

In the nonlinear regime, the electric field produces a partial filling of the Coulomb gap. 
This effect can be explained in terms of the effective temperature. In equilibrium, the density of states at the Fermi Level, $g(0)$, depends
on $T$ and in 2D this dependence is linear at low temperatures. We previously showed that  in relaxation experiments $g(0)$ as a function of $T_{\rm eff}$ presents the same dependence as in equilibrium \cite{SoOr08}. 
Here we found that the same relationship holds in the non linear regime. 
In conclusion, the quantity $g(0)$, which could be measured experimentally, constitutes an excellent thermometer for the electrons effective temperature.

\section{Nonlinear conductivity}

\begin{figure}[htb]
\includegraphics*[width=.45\textwidth]{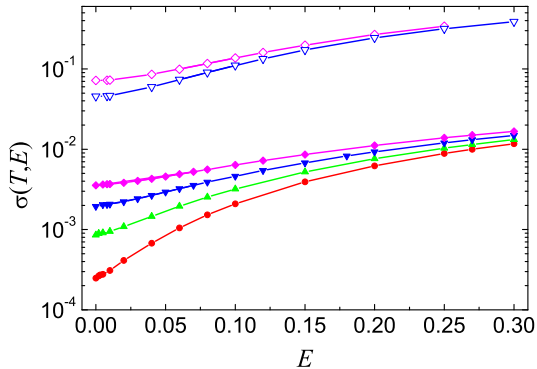}
\caption{%
(Color online) Conductivity on a logarithmic scale as a function of $E$ for $T=0.06$ (red circles), 0.08 (green up triangles), 0.1 (blue down triangles) and 0.12 (magenta diamonds), and 
$\xi=1$ (solid symbols) and $2$ (empty symbols). } \label{fig5}
\end{figure}

We have calculated the conductivity for different values of the applied field and
the temperature in the variable range hopping regime.
We observe a very small linear regime for small values of the 
field and then a systematic increase of the conductivity
for high values of $E$. This increase also depends on $T$. In Fig.\ \ref{fig5} we represent the conductivity
$\sigma(T,E)$ on a logarithmic
scale as a function of  $E$ for the same values of $T$ as in Fig.\ \ref{fig3}. 
The temperatures are $T=0.06$ (circles), 0.08 (up triangles), 0.1 (down triangles) and 0.12 (diamonds), and the localizations lengths are
$\xi=1$ (solid symbols) and $2$ (empty symbols). The lines are a guide to the eye.
For high field values, the temperature dependence of $\sigma(T,E)$ is weak.
For our model, the conductivity tends to zero for very high values of $E$ \cite{CaVo09} (not shown in the figure).
Later on we will analyze how to collapse on a single curve the data for different values of $T$ and the applicability of the so-called field-effect models and the hot-electron model.

\begin{figure}[htb]
\includegraphics*[width=.45\textwidth]{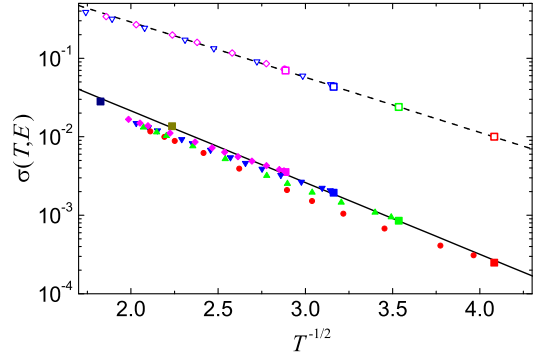}
\caption{%
(Color online) $\sigma(T,0)$ on a logarithmic scale  as a function of  $T^{-1/2}$ (large squares) for $\xi=1$ (solid symbols, lower set of points) and 2 (empty symbols, upper set).
We also plot $\sigma(T,E)$ in  the nonlinear regime as a function this time of $T_{\rm eff}^{-1/2}$.
The data and the symbols employed in the non linear regime are the same as in Fig.\ \ref{fig5}.} \label{fig7}
\end{figure}

The conductivity in the linear regime follows the $T^{-1/2}$ law as can be seen
in Fig.\ \ref{fig7}, where $\sigma(T,0)$ is represented by large squares on a logarithmic scale as a function of  $T^{-1/2}$. The lower set of data (solid symbols) correspond to $\xi=1$ and the upper set (empty symbols) to $\xi=2$. 
The straight lines are fit to these linear regime data.
The relatively good fitting of the data to a straight
line indicates that the standard form for variable range hopping in the presence of interactions, Eq.\ (\ref{unmedio}), is fairly well satisfied.
An slightly better linear fit is obtained if we represent $\sigma(T,0)T$, instead of $\sigma(T,0)$,
avoiding certain logarithmic corrections.
The characteristic temperature $T_0$ would also be in that case in better
agreement with  previous results \cite{TsEf02}.
However, we prefer to represent $\sigma(T,0)$ itself to keep the strongest possible analogy with the usual analysis of experimental results in this area and to avoid having to rescale the vertical axis with $T_{\rm eff}$. Our conclusions do not depend on this choice. 
From the slope of the straight lines we found that the characteristic
temperature $T_0$ is in this case equal to 4.43 for $\xi=1$ and 2.62 for $\xi=2$.
We found that the constant of proportionality implicit in Eq.\ (\ref{unmedio}) slightly depends on $\xi$.

As we can determine $T_{\rm eff}$ directly from the long one-electron excitation probabilities, we can check the
validity of the hot-electron model, which assumes that at a given $E$ and $T$ the conductivity
is the same as the linear conductivity at the corresponding $T_{\rm eff}$, i.e., $\sigma(T,E)=\sigma(T_{\rm eff},0)$. 
In Fig.\ \ref{fig7}, we have also represented $\sigma(T,E)$ on a logarithmic scale
as a function of the corresponding $T_{\rm eff}^{-1/2}$ for $T=0.06$ (circles), 0.08 (up triangles), 0.1 (down triangles) and 0.12 (diamonds). The solid symbols correspond to $\xi=1$ and the empty symbols to $2$. 
We see that the data in the nonlinear regime lie relatively close to the  $T^{-1/2}$ law corresponding to the linear regime. The difference between $\sigma(T,E)$ and $\sigma(T_{\rm eff},0)$ is at most a factor of two. 
The quality of the agreement of the data for  $\xi=2$ with the hot-electron model  is quite good, much better than for the case $\xi=1$. This trend is in line with experimental results
that show that the hot-electron model works better for lower values of $T_0/T$, which in our simulations is equivalent to larger values of $\xi$.
The increase of $\sigma$ as a function of the electric field can be roughly interpreted as due to an increase in $T_{\rm eff}$. 
The energy factor entering in the transition rates between two configurations can be decomposed 
into two terms, one due to the occupation of the configurations and the other to the energy penalty of the transition. The first one depends on the effective temperature, while the second one depends, in principle, on the phonon temperature $T$. Fig.\ \ref{fig7} indicates that the contribution of the first term masks to a large degree the effect of the second term.

In order to further check the predictions of the hot-electron model we must study the power
dissipated by the system.

\section{Dissipated power}

The relation between the effective temperature and the electric field must be linked to the 
energy dissipation capability of the system \cite{MaFl91,AlKr09}. 
Most authors have assumed that the power supplied by the electric field is related to
$T_{\rm eff}$ through Eq.\ (\ref{power}). This type of expression was obtained for metals, and it was also used, with different values of the exponent $\beta$, in the the diffusive
regime and even in the localized regime without any theoretical support.
$T_{\rm eff}$ is usually obtained from the hot-electron assumption, that is,
through the relation $\sigma(T,E)=\sigma(T_{\rm eff},0)$.
In the cases where the hot-electron model works well, the power provided by the electric field is indeed approximately related to the effective temperature by Eq.\ (\ref{power}).

In our simulations, we can obtain more detailed information than in experiments since we can measure $T_{\rm eff}$ directly
from the excitation spectra, and because we can compute 
separately the powers absorbed and emitted by the electrons.

\begin{figure}[htb]
\includegraphics*[width=.45\textwidth]{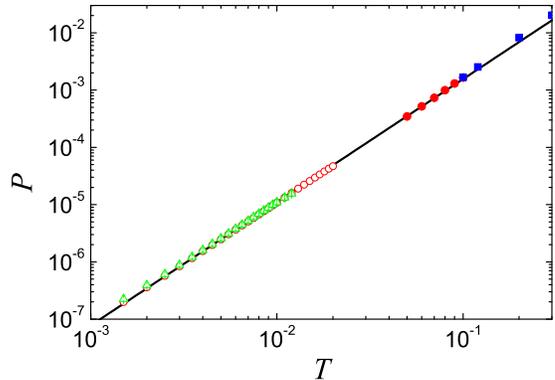}
\caption{%
(Color online) Absorbed power in equilibrium as a function of $T$ obtained with Monte Carlo simulations (solid symbols) and through 
a set of low energy configurations (empty symbols). Different symbols correspond to different system sizes: $N=500$ (green triangles), 1000 (red circles) and 2000 (black squares). } \label{fig5a}
\end{figure}

We have first calculated the power absorbed by the electronic system from the phonons at equilibrium at a temperature $T$ in the absence of an electric field (in this case, this power is of course equal to the power emitted back by the electrons). In Fig.\ \ref{fig5a} we show the absorbed power in equilibrium as a function of $T$. This power has been obtained by two different methods: Monte Carlo simulations (solid symbols) and through 
a set of low energy configurations (empty symbols). Different symbols correspond to different system sizes. In the range $0.001<T<0.3$, we found that the data can be fitted by
$bT^\gamma$, with $\gamma=2.151\pm 0.001$, independent of the localization length $\xi$, and $b$ a constant weakly dependent on $\xi$. The value of $\gamma$ is much smaller than
the values of the exponents found in the expressions for the dissipated power,  usually between 5 and 6 (see, for example, Ref.\ \onlinecite{WaWe90}). This discrepancy is due to the fact that most of the exchange of energy in equilibrium is  absorbed and emitted by short excitations, contributing as $T^2$, while most of the dissipated power in DC current is related to more complex excitations. 

\begin{figure}[htb]
\includegraphics*[width=.45\textwidth]{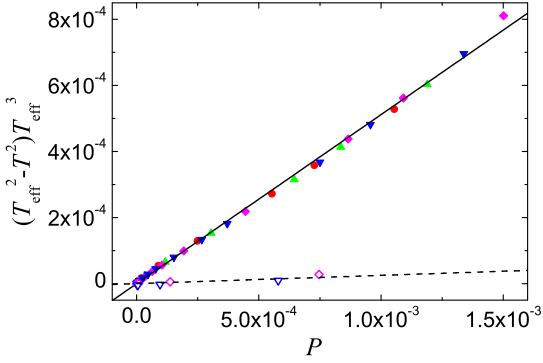}
\caption{%
(Color online) $\left(T_{\rm eff}^2-T^2\right)T_{\rm eff}^3$ versus $P$ for
$T=0.06$ (red circles), 0.08 (green up triangles), 0.1 (blue down triangles) and 0.12 (magenta diamonds), and $\xi=1$ (solid symbols) and 2 (empty symbols). The straight lines are a linear fit passing through the origin for $\xi=1$ (continuous) and 2 (dashed). } \label{fig8}
\end{figure}

We now apply an electric field and measure the absorbed and emitted powers in terms of
$T$ and $E$. The difference between them is, of course, the electric power provided by the field  $\sigma(T,E)E^2$. We observe that the power emitted by the system is equal to the value at equilibrium plus a (usually smaller) term that only depends on $T_{\rm eff}$ and can be pretty well represented by a power law, with an exponent close to 5. However, the absorbed power is equal to the power in equilibrium plus a term that depends on both $T$ and $T_{\rm eff}$.
This is in disagreement with the standard assumption, i.e., that it is a function of $T$ only.
We found that the  power provided by the electric field can be fitted pretty well by an expression of the form
\begin{equation}
P=\sigma(T,E) E^2=c\left(T_{\rm eff}^\alpha-T^\alpha\right)T_{\rm eff}^{\beta-\alpha}.
\label{dis}\end{equation}
The best fit corresponds to $\alpha=1.99$ and $\beta=5.26$.
We have considered $\alpha=2$ and $\beta=5$ for simplicity and because the fit is also good for these exponents.
We note that the right-hand side in Eq.\ (\ref{dis}) is the difference between the powers emitted and absorbed by the electrons, but each term separately does not correspond to the full emitted and absorbed powers, since the large contribution by short dipoles is equal for both powers and cancels.
In Fig.\ \ref{fig8} we represent  $\left(T_{\rm eff}^2-T^2\right)T_{\rm eff}^3$ as a function of the dissipated power $P$ for several temperatures, $T=0.06$ (circles), 0.08 (up triangles), 0.1 (down triangles) and 0.12 (diamonds). Solid symbols correspond to a localization length $\xi=1$ and empty symbols to $\xi=2$. 
The straight lines are  linear fit to the data, forced to pass through the origin,
for $\xi=1$ (continuous line) and 2 (dashed line).
We note both the good overlap of the data and the global linear behavior.
The proportionality constant $c$ depends strongly on $\xi$.

\begin{figure}[htb]
\includegraphics*[width=.45\textwidth]{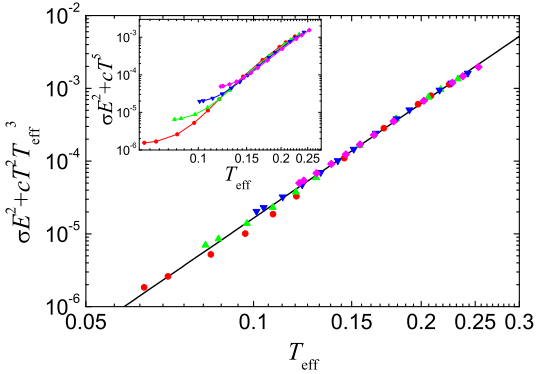}
\caption{%
(Color online) $\sigma(T,E)E^2+cT^2T_{\rm eff}^3$ as a function of $T_{\rm eff}$ for the same data as in Fig.\ \ref{fig8}. Inset: $\sigma(T,E)E^2+cT^5$ as a function of $T_{\rm eff}$ for the same data as in the main part.} \label{fig9}
\end{figure}

An alternative and more demanding way to represent the data in figure \ref{fig8}, usually employed in the literature \cite{WaWe90}, is by plotting the electric power plus the $T$ dependent contribution in Eq.\ (\ref{dis}) (the extra absorbed power) as a function of $T_{\rm eff}$. 
In Fig.\ \ref{fig9} we plot $\sigma(T,E)E^2+cT^2T_{\rm eff}^3$, where $c$ is the inverse of the slope of the straight line in Fig.\ \ref{fig8}, versus $T_{\rm eff}$ on a double logarithmic scale. We again see a good overlap 
of the data and a fairly linear behavior, indicating in this case a good power law dependence of the emitted power on $T_{\rm eff}$.
The straight line is a linear fit to the data and its slope is $\beta=5.25$. This  value of $\beta$ is the same as our best estimate for $\beta$ obtained from an overall nonlinear fitting, which is not surprising
since the increase in the emitted power with respect to equilibrium is much larger than the increase in the absorbed power.

In the inset of Fig.\ \ref{fig9} we represent $\sigma(T,E)E^2+cT^5$ as a function of $T_{\rm eff}$ for the same data as in the main part of the figure. It clearly shows that  the data for small electric fields
deviate from a straight line and do not overlap for different temperatures. A similar behavior is observed in several experimental results \cite{WaWe90}. Thus, we conclude that Eq.\ 
(\ref{dis}) for the dissipated power is better than Eq.\ (\ref{power}), usually employed in the literature.
The use of Eq.\ (\ref{dis}), instead of Eq.\ (\ref{power}), can explain many ``anomalies" found in experiments for low dissipated powers as, for example, in two-dimensional interfaces \cite{MaFl91,MiSh04},
in neutron-transmutation-doped Ge \cite{WaWe90} and in amorphous indium oxide films in the insulating regime \cite{OvSa09}.

The values of the exponents $\alpha$ and $\beta$, entering in Eq.\ (\ref{dis}), found in our simulations depend on the model used and it may be different in experimental situations, depending for example on the density of phonons and on the dimensionality of the system. 
The corresponding exponents may depend on the particular situation studied and remain to be measured experimentally or properly calculated. 
The important message is that the absorbed power depends on both $T$ and $T_{\rm eff}$ and the dissipated power is given by Eq.\ (\ref{dis}), instead of 
Eq.\ (\ref{power}).

\section{Hot-electron model}

We have combined Eqs.\ (\ref{unmedioeff}) and (\ref{dis}) to obtained our modified version of the hot-electron model. The value of $T_0$ and the proportionality constant in Eq.\ (\ref{unmedioeff}) are empirically obtained from the fit of the data in the linear regime in Fig.\ \ref{fig7}. In Eq.\ (\ref{dis}) we take $\alpha=2$ and $\beta=5$ and the constant $c$ is obtained from Fig.\ \ref{fig8}. With these values of the parameters we find numerically $T_{\rm eff}$ and $\sigma(T,E)$ for each value of $T$ and $E$. 
We note that there are many different implementations of the hot-electron model in the literature and the one more often found in experimental papers is not the same as ours.
The latter obtains $T_{\rm eff}$ empirically from the relation $\sigma(T,E)=\sigma(T_{\rm eff},0)$ and uses this value to fit the expression of the power. 

\begin{figure}[htb]
\includegraphics*[width=.45\textwidth]{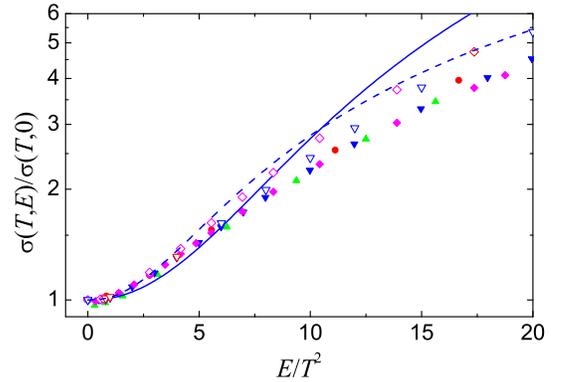}
\caption{%
(Color online) $\sigma(T,E)/\sigma(T,0)$, on a logarithmic scale, as a function of
 ${E}/T^2$ for the same data as in Fig.\ \ref{fig5}. The curves correspond to the predictions of the hot-electron model for $\xi=1$ (continuous line) and 2 (dashed line).} \label{fig6}
\end{figure}

We found empirically that the values of $\sigma(T,E)$ for different temperatures and electric fields can be overlapped relatively well  
by plotting $\sigma(T,E)/\sigma(T,0)$ as a function of $E/T^2$.
In Fig.\ \ref{fig6} we plot $\sigma(T,E)/\sigma(T,0)$ (symbols) on a logarithmic scale  versus  $E/T^2$ for the same data as in Fig.\ \ref{fig5}. 
The curves correspond to the predictions of the hot-electron model  for $\xi=1$ (continuous line) and 2 (dashed line) and $T=0.1$. The predictions for other values of $T$ overlap between themselves relatively well for each $\xi$.  The agreement between the hot-electron model and our simulations is reasonable, specially for $\xi=2$.

According to field effect models, Eq.\ (\ref{field}), $\log(\sigma(T,E)/\sigma(T,0))$ should be proportional to $EL/T$. We found that this dependence is not fully verified,
but there is no strong disagreement either.
The collapse of the conductivity data as a function of $E/T^2$ would indicate that the characteristic hopping length for nonlinear effects would be inversely proportional to $T$ as predicted by Shklovskii's model. However we note  that the localization length dependence is not the one expected.

\section{Conclusions}

We observed that slow electronic modes of CG are thermalized at an effective temperature higher than $T$ in the non linear regime. 
Although this effective temperature is well defined, as we verified through the occupation distribution, the density of long excitations and the fluctuation-dissipation theorem,
the detail mechanism underlying this thermalization is not clear and deserves farther investigation.
For moderate electric fields, the conductivity increases with the field and this increment can be roughly interpreted in terms of the increase in the effective temperature.

The non linear conductivity in terms of the effective temperature roughly follows the $T^{-1/2}$ law with a characteristic temperature $T_0$ similar to that of the linear regime.
From a theoretical analysis one could conclude that this $T_0$ should be different from that of the linear regime and even may depend on $T$. According to our results these deviations from the linear regime are small.

The power provided by the electric field in the strongly localized systems that we consider can be expressed as a difference of powers of $T_{\rm eff}$ and $T$. However, the expression that fits better the numerical results, Eq.\ (\ref{dis}), is not the one usually employed in the literature. It would be interesting to find the exponent $\alpha$ and $\beta$ from experimental results.

The switching transition observed in an amorphous indium oxide film in the insulating state \cite{OvSa09} was explained in terms of a bistability of the electron temperature $T_{\rm eff}$, as obtained from  the hot-electron model \cite{AlKr09}.
This effect might be observed in CG in the variable-range hopping regime with $T_0/T\ll 1$ or in another hopping regime with a stronger temperature dependence of the conductivity.
The possible mechanism for switching in our system is related to the slow degrees of freedom. The current heats the electronic system and part of this heat remains in the slow
modes producing an effective temperature, which strongly increases conductivity.
 In fact Eq.\ (\ref{dis}) for the power, with $\alpha=2$ and $\beta=5$, and Eq.\ (\ref{unmedioeff}) for the conductivity, predict that the bistability will occur at temperatures lower than $T_0/(144\sqrt{3})$ and in this case the hot-electron model should work pretty well. This temperature range is difficult to  simulate with our present MC method due to the glassy nature of the system, but this is an interesting problem that deserves further study. 

\acknowledgments

We thanks useful discussions with Michael Pollak, Joakim Bergli, Zvi Ovadyahu and
Igor Lerner. 
We also acknowledge financial support from 
projects FIS2009-13483 (MICINN) and 08832/PI/08 (Fundacion Seneca).


\begin{thebibliography}{34}
\expandafter\ifx\csname natexlab\endcsname\relax\def\natexlab#1{#1}\fi
\expandafter\ifx\csname bibnamefont\endcsname\relax
  \def\bibnamefont#1{#1}\fi
\expandafter\ifx\csname bibfnamefont\endcsname\relax
  \def\bibfnamefont#1{#1}\fi
\expandafter\ifx\csname citenamefont\endcsname\relax
  \def\citenamefont#1{#1}\fi
\expandafter\ifx\csname url\endcsname\relax
  \def\url#1{\texttt{#1}}\fi
\expandafter\ifx\csname urlprefix\endcsname\relax\def\urlprefix{URL }\fi
\providecommand{\bibinfo}[2]{#2}
\providecommand{\eprint}[2][]{\url{#2}}

\bibitem[{\citenamefont{Mott}(1968)}]{Mo68}
\bibinfo{author}{\bibfnamefont{N.~F.} \bibnamefont{Mott}},
  \bibinfo{journal}{Journal of Non-Crystalline Solids}
  \textbf{\bibinfo{volume}{1}}, \bibinfo{pages}{1} (\bibinfo{year}{1968}).

\bibitem[{\citenamefont{Efros and Shklovskii}(1975)}]{EfSh75}
\bibinfo{author}{\bibfnamefont{A.~L.} \bibnamefont{Efros}} \bibnamefont{and}
  \bibinfo{author}{\bibfnamefont{B.~I.} \bibnamefont{Shklovskii}},
  \bibinfo{journal}{Journal of Physics C} \textbf{\bibinfo{volume}{8}},
  \bibinfo{pages}{L49} (\bibinfo{year}{1975}).

\bibitem[{\citenamefont{Levin et~al.}(1987)\citenamefont{Levin, Nguyen,
  Shklovskii, and Efros}}]{LeNg87}
\bibinfo{author}{\bibfnamefont{E.~I.} \bibnamefont{Levin}},
  \bibinfo{author}{\bibfnamefont{V.~L.} \bibnamefont{Nguyen}},
  \bibinfo{author}{\bibfnamefont{B.~I.} \bibnamefont{Shklovskii}},
  \bibnamefont{and} \bibinfo{author}{\bibfnamefont{A.~L.} \bibnamefont{Efros}},
  \bibinfo{journal}{Soviet Physics JETP} \textbf{\bibinfo{volume}{65}},
  \bibinfo{pages}{842} (\bibinfo{year}{1987}).

\bibitem[{\citenamefont{Pollak and Riess}(1976)}]{PoRi76}
\bibinfo{author}{\bibfnamefont{M.}~\bibnamefont{Pollak}} \bibnamefont{and}
  \bibinfo{author}{\bibfnamefont{I.}~\bibnamefont{Riess}},
  \bibinfo{journal}{Journal of Physics C-Solid State Physics}
  \textbf{\bibinfo{volume}{9}}, \bibinfo{pages}{2339} (\bibinfo{year}{1976}).

\bibitem[{\citenamefont{Shklovskii}(1976)}]{Sh76}
\bibinfo{author}{\bibfnamefont{B.~I.} \bibnamefont{Shklovskii}},
  \bibinfo{journal}{Soviet Physics Semiconductors}
  \textbf{\bibinfo{volume}{10}}, \bibinfo{pages}{855} (\bibinfo{year}{1976}).

\bibitem[{\citenamefont{Grannan et~al.}(1992)\citenamefont{Grannan, Lange,
  Haller, and Beeman}}]{GrLa92}
\bibinfo{author}{\bibfnamefont{S.~M.} \bibnamefont{Grannan}},
  \bibinfo{author}{\bibfnamefont{A.~E.} \bibnamefont{Lange}},
  \bibinfo{author}{\bibfnamefont{E.~E.} \bibnamefont{Haller}},
  \bibnamefont{and} \bibinfo{author}{\bibfnamefont{J.~W.}
  \bibnamefont{Beeman}}, \bibinfo{journal}{Physical Review B}
  \textbf{\bibinfo{volume}{45}}, \bibinfo{pages}{4516} (\bibinfo{year}{1992}).

\bibitem[{\citenamefont{Zhang et~al.}(1998)\citenamefont{Zhang, Cui, Juda,
  McCammon, Kelley, Moseley, Stahle, and Szymkowiak}}]{ZhCu98a}
\bibinfo{author}{\bibfnamefont{J.}~\bibnamefont{Zhang}},
  \bibinfo{author}{\bibfnamefont{W.}~\bibnamefont{Cui}},
  \bibinfo{author}{\bibfnamefont{M.}~\bibnamefont{Juda}},
  \bibinfo{author}{\bibfnamefont{D.}~\bibnamefont{McCammon}},
  \bibinfo{author}{\bibfnamefont{R.~L.} \bibnamefont{Kelley}},
  \bibinfo{author}{\bibfnamefont{S.~H.} \bibnamefont{Moseley}},
  \bibinfo{author}{\bibfnamefont{C.~K.} \bibnamefont{Stahle}},
  \bibnamefont{and} \bibinfo{author}{\bibfnamefont{A.~E.}
  \bibnamefont{Szymkowiak}}, \bibinfo{journal}{Physical Review B}
  \textbf{\bibinfo{volume}{57}}, \bibinfo{pages}{4472} (\bibinfo{year}{1998}).

\bibitem[{\citenamefont{Wang et~al.}(1990)\citenamefont{Wang, Wellstood,
  Sadoulet, Haller, and Beeman}}]{WaWe90}
\bibinfo{author}{\bibfnamefont{N.}~\bibnamefont{Wang}},
  \bibinfo{author}{\bibfnamefont{F.~C.} \bibnamefont{Wellstood}},
  \bibinfo{author}{\bibfnamefont{B.}~\bibnamefont{Sadoulet}},
  \bibinfo{author}{\bibfnamefont{E.~E.} \bibnamefont{Haller}},
  \bibnamefont{and} \bibinfo{author}{\bibfnamefont{J.}~\bibnamefont{Beeman}},
  \bibinfo{journal}{Physical Review B} \textbf{\bibinfo{volume}{41}},
  \bibinfo{pages}{3761} (\bibinfo{year}{1990}).

\bibitem[{\citenamefont{Gershenson et~al.}(2000)\citenamefont{Gershenson,
  Khavin, Reuter, Schafmeister, and Wieck}}]{GeKh00}
\bibinfo{author}{\bibfnamefont{M.~E.} \bibnamefont{Gershenson}},
  \bibinfo{author}{\bibfnamefont{Y.~B.} \bibnamefont{Khavin}},
  \bibinfo{author}{\bibfnamefont{D.}~\bibnamefont{Reuter}},
  \bibinfo{author}{\bibfnamefont{P.}~\bibnamefont{Schafmeister}},
  \bibnamefont{and} \bibinfo{author}{\bibfnamefont{A.~D.} \bibnamefont{Wieck}},
  \bibinfo{journal}{Physical Review Letters} \textbf{\bibinfo{volume}{85}},
  \bibinfo{pages}{1718} (\bibinfo{year}{2000}).

\bibitem[{\citenamefont{Minkov et~al.}(2004)\citenamefont{Minkov, Sherstobitov,
  Rut, and Germanenko}}]{MiSh04}
\bibinfo{author}{\bibfnamefont{G.~M.} \bibnamefont{Minkov}},
  \bibinfo{author}{\bibfnamefont{A.}~\bibnamefont{Sherstobitov}},
  \bibinfo{author}{\bibfnamefont{O.}~\bibnamefont{Rut}}, \bibnamefont{and}
  \bibinfo{author}{\bibfnamefont{A.}~\bibnamefont{Germanenko}},
  \bibinfo{journal}{Physica E} \textbf{\bibinfo{volume}{25}},
  \bibinfo{pages}{42} (\bibinfo{year}{2004}).

\bibitem[{\citenamefont{Jain and Raychaudhuri}(2008)}]{JaRa08}
\bibinfo{author}{\bibfnamefont{H.}~\bibnamefont{Jain}} \bibnamefont{and}
  \bibinfo{author}{\bibfnamefont{A.~K.} \bibnamefont{Raychaudhuri}},
  \bibinfo{journal}{Applied Physics Letters} \textbf{\bibinfo{volume}{93}},
  \bibinfo{pages}{182110} (\bibinfo{year}{2008}).

\bibitem[{\citenamefont{Fisher et~al.}(2009)\citenamefont{Fisher, Genossar,
  Patlagan, and Reisner}}]{FiGe09}
\bibinfo{author}{\bibfnamefont{B.}~\bibnamefont{Fisher}},
  \bibinfo{author}{\bibfnamefont{J.}~\bibnamefont{Genossar}},
  \bibinfo{author}{\bibfnamefont{L.}~\bibnamefont{Patlagan}}, \bibnamefont{and}
  \bibinfo{author}{\bibfnamefont{G.~M.} \bibnamefont{Reisner}},
  \bibinfo{journal}{Applied Physics Letters} \textbf{\bibinfo{volume}{95}},
  \bibinfo{pages}{132501} (\bibinfo{year}{2009}).

\bibitem[{\citenamefont{Galeazzi et~al.}(2007)\citenamefont{Galeazzi, Liu,
  McCammon, Rocks, Sanders, Smith, Tan, Vaillancourt, Boyce, Brekosky
  et~al.}}]{GaLi07}
\bibinfo{author}{\bibfnamefont{M.}~\bibnamefont{Galeazzi}},
  \bibinfo{author}{\bibfnamefont{D.}~\bibnamefont{Liu}},
  \bibinfo{author}{\bibfnamefont{D.}~\bibnamefont{McCammon}},
  \bibinfo{author}{\bibfnamefont{L.~E.} \bibnamefont{Rocks}},
  \bibinfo{author}{\bibfnamefont{W.~T.} \bibnamefont{Sanders}},
  \bibinfo{author}{\bibfnamefont{B.}~\bibnamefont{Smith}},
  \bibinfo{author}{\bibfnamefont{P.}~\bibnamefont{Tan}},
  \bibinfo{author}{\bibfnamefont{J.~E.} \bibnamefont{Vaillancourt}},
  \bibinfo{author}{\bibfnamefont{K.~R.} \bibnamefont{Boyce}},
  \bibinfo{author}{\bibfnamefont{R.~P.} \bibnamefont{Brekosky}},
  \bibnamefont{et~al.}, \bibinfo{journal}{Physical Review B}
  \textbf{\bibinfo{volume}{76}}, \bibinfo{pages}{155207}
  (\bibinfo{year}{2007}).

\bibitem[{\citenamefont{Leturcq et~al.}(2003)\citenamefont{Leturcq, L'H\^{o}te,
  Tourbot, Senz, Gennser, Ihn, Ensslin, Dehlinger, and Gr\"utzmacher}}]{LeLh03}
\bibinfo{author}{\bibfnamefont{R.}~\bibnamefont{Leturcq}},
  \bibinfo{author}{\bibfnamefont{D.}~\bibnamefont{L'H\^{o}te}},
  \bibinfo{author}{\bibfnamefont{R.}~\bibnamefont{Tourbot}},
  \bibinfo{author}{\bibfnamefont{V.}~\bibnamefont{Senz}},
  \bibinfo{author}{\bibfnamefont{U.}~\bibnamefont{Gennser}},
  \bibinfo{author}{\bibfnamefont{T.}~\bibnamefont{Ihn}},
  \bibinfo{author}{\bibfnamefont{K.}~\bibnamefont{Ensslin}},
  \bibinfo{author}{\bibfnamefont{G.}~\bibnamefont{Dehlinger}},
  \bibnamefont{and}
  \bibinfo{author}{\bibfnamefont{D.}~\bibnamefont{Gr\"utzmacher}},
  \bibinfo{journal}{Europhysics Letters} \textbf{\bibinfo{volume}{61}},
  \bibinfo{pages}{499–505} (\bibinfo{year}{2003}).

\bibitem[{\citenamefont{Yu et~al.}(2004)\citenamefont{Yu, Wang, Wehrenberg, and
  Guyot-Sionnest}}]{YuWa04}
\bibinfo{author}{\bibfnamefont{D.}~\bibnamefont{Yu}},
  \bibinfo{author}{\bibfnamefont{C.~J.} \bibnamefont{Wang}},
  \bibinfo{author}{\bibfnamefont{B.~L.} \bibnamefont{Wehrenberg}},
  \bibnamefont{and}
  \bibinfo{author}{\bibfnamefont{P.}~\bibnamefont{Guyot-Sionnest}},
  \bibinfo{journal}{Physical Review Letters} \textbf{\bibinfo{volume}{92}},
  \bibinfo{pages}{216802} (\bibinfo{year}{2004}).

\bibitem[{\citenamefont{Ladieu et~al.}(2000)\citenamefont{Ladieu, L'H\^{o}te,
  and Tourbot}}]{LaLh00}
\bibinfo{author}{\bibfnamefont{F.}~\bibnamefont{Ladieu}},
  \bibinfo{author}{\bibfnamefont{D.}~\bibnamefont{L'H\^{o}te}},
  \bibnamefont{and} \bibinfo{author}{\bibfnamefont{R.}~\bibnamefont{Tourbot}},
  \bibinfo{journal}{Physical Review B} \textbf{\bibinfo{volume}{61}},
  \bibinfo{pages}{8108} (\bibinfo{year}{2000}).

\bibitem[{\citenamefont{Ovadia et~al.}(2009)\citenamefont{Ovadia, Sac\'ep\'e,
  and Shahar}}]{OvSa09}
\bibinfo{author}{\bibfnamefont{M.}~\bibnamefont{Ovadia}},
  \bibinfo{author}{\bibfnamefont{B.}~\bibnamefont{Sac\'ep\'e}},
  \bibnamefont{and} \bibinfo{author}{\bibfnamefont{D.}~\bibnamefont{Shahar}},
  \bibinfo{journal}{Physical Review Letters} \textbf{\bibinfo{volume}{102}},
  \bibinfo{pages}{176802} (\bibinfo{year}{2009}).

\bibitem[{\citenamefont{Altshuler et~al.}(2009)\citenamefont{Altshuler,
  Kravtsov, Lerner, and Aleiner}}]{AlKr09}
\bibinfo{author}{\bibfnamefont{B.~L.} \bibnamefont{Altshuler}},
  \bibinfo{author}{\bibfnamefont{V.~E.} \bibnamefont{Kravtsov}},
  \bibinfo{author}{\bibfnamefont{I.~V.} \bibnamefont{Lerner}},
  \bibnamefont{and} \bibinfo{author}{\bibfnamefont{I.~L.}
  \bibnamefont{Aleiner}}, \bibinfo{journal}{Physical Review Letters}
  \textbf{\bibinfo{volume}{102}}, \bibinfo{pages}{176803}
  (\bibinfo{year}{2009}).

\bibitem[{\citenamefont{Ladieu et~al.}(1996)\citenamefont{Ladieu, Sanquer, and
  Bouchaud}}]{LaSa96}
\bibinfo{author}{\bibfnamefont{F.}~\bibnamefont{Ladieu}},
  \bibinfo{author}{\bibfnamefont{M.}~\bibnamefont{Sanquer}}, \bibnamefont{and}
  \bibinfo{author}{\bibfnamefont{J.~P.} \bibnamefont{Bouchaud}},
  \bibinfo{journal}{Physical Review B} \textbf{\bibinfo{volume}{53}},
  \bibinfo{pages}{973} (\bibinfo{year}{1996}).

\bibitem[{\citenamefont{Somoza et~al.}(2008)\citenamefont{Somoza, Ortu\~no,
  Caravaca, and Pollak}}]{SoOr08}
\bibinfo{author}{\bibfnamefont{A.~M.} \bibnamefont{Somoza}},
  \bibinfo{author}{\bibfnamefont{M.}~\bibnamefont{Ortu\~no}},
  \bibinfo{author}{\bibfnamefont{M.}~\bibnamefont{Caravaca}}, \bibnamefont{and}
  \bibinfo{author}{\bibfnamefont{M.}~\bibnamefont{Pollak}},
  \bibinfo{journal}{Physical Review Letters} \textbf{\bibinfo{volume}{101}},
  \bibinfo{pages}{056601} (\bibinfo{year}{2008}).

\bibitem[{\citenamefont{Pollak and Ortu\~no}(1985)}]{PoOr85}
\bibinfo{author}{\bibfnamefont{M.}~\bibnamefont{Pollak}} \bibnamefont{and}
  \bibinfo{author}{\bibfnamefont{M.}~\bibnamefont{Ortu\~no}},
  \emph{\bibinfo{title}{Electron-electron interactions in disordered systems}}
  (\bibinfo{publisher}{North-Holland}, \bibinfo{year}{1985}), chap.
  \bibinfo{chapter}{The effect of Coulomb interactions on electronic states and
  transport in disordered insulators}, pp. \bibinfo{pages}{287--408}.

\bibitem[{\citenamefont{Davies et~al.}(1984)\citenamefont{Davies, Lee, and
  Rice}}]{DaLe84}
\bibinfo{author}{\bibfnamefont{J.~H.} \bibnamefont{Davies}},
  \bibinfo{author}{\bibfnamefont{P.~A.} \bibnamefont{Lee}}, \bibnamefont{and}
  \bibinfo{author}{\bibfnamefont{T.~M.} \bibnamefont{Rice}},
  \bibinfo{journal}{Physical Review B} \textbf{\bibinfo{volume}{29}},
  \bibinfo{pages}{4260} (\bibinfo{year}{1984}).

\bibitem[{\citenamefont{Somoza et~al.}(2006)\citenamefont{Somoza, Ortu\~no, and
  Pollak}}]{SoOr06}
\bibinfo{author}{\bibfnamefont{A.~M.} \bibnamefont{Somoza}},
  \bibinfo{author}{\bibfnamefont{M.}~\bibnamefont{Ortu\~no}}, \bibnamefont{and}
  \bibinfo{author}{\bibfnamefont{M.}~\bibnamefont{Pollak}},
  \bibinfo{journal}{Physical Review B} \textbf{\bibinfo{volume}{73}},
  \bibinfo{pages}{045123} (\bibinfo{year}{2006}).

\bibitem[{\citenamefont{Somoza and Ortu\~no}(2005)}]{SoOr05}
\bibinfo{author}{\bibfnamefont{A.~M.} \bibnamefont{Somoza}} \bibnamefont{and}
  \bibinfo{author}{\bibfnamefont{M.}~\bibnamefont{Ortu\~no}},
  \bibinfo{journal}{Physical Review B} \textbf{\bibinfo{volume}{72}},
  \bibinfo{pages}{224202} (\bibinfo{year}{2005}).

\bibitem{KwLa04}
\bibinfo{author}{\bibfnamefont{W.} \bibnamefont{Kwak}},
  \bibinfo{author}{\bibfnamefont{D.~P.}~\bibnamefont{Landau}}, \bibnamefont{and}
  \bibinfo{author}{\bibfnamefont{B.}~\bibnamefont{Schmittmann}},
  \bibinfo{journal}{Physical Review E} \textbf{\bibinfo{volume}{69}},
  \bibinfo{pages}{066134} (\bibinfo{year}{2004}).

\bibitem[{\citenamefont{Tsigankov et~al.}(2003)\citenamefont{Tsigankov, Pazy,
  Laikhtman, and Efros}}]{TsPa03}
\bibinfo{author}{\bibfnamefont{D.~N.} \bibnamefont{Tsigankov}},
  \bibinfo{author}{\bibfnamefont{E.}~\bibnamefont{Pazy}},
  \bibinfo{author}{\bibfnamefont{B.~D.} \bibnamefont{Laikhtman}},
  \bibnamefont{and} \bibinfo{author}{\bibfnamefont{A.~L.} \bibnamefont{Efros}},
  \bibinfo{journal}{Physical Review B} \textbf{\bibinfo{volume}{68}},
  \bibinfo{pages}{184205} (\bibinfo{year}{2003}).

\bibitem[{\citenamefont{Bortz et~al.}(1975)\citenamefont{Bortz, Kalos, and
  Lebowitz}}]{BoKa75}
\bibinfo{author}{\bibfnamefont{A.~B.} \bibnamefont{Bortz}},
  \bibinfo{author}{\bibfnamefont{M.~H.} \bibnamefont{Kalos}}, \bibnamefont{and}
  \bibinfo{author}{\bibfnamefont{J.~L.} \bibnamefont{Lebowitz}},
  \bibinfo{journal}{Journal of Computational Physics}
  \textbf{\bibinfo{volume}{17}}, \bibinfo{pages}{10} (\bibinfo{year}{1975}).

\bibitem[{\citenamefont{M\"obius and Thomas}(1997)}]{MoTh97}
\bibinfo{author}{\bibfnamefont{A.}~\bibnamefont{M\"obius}} \bibnamefont{and}
  \bibinfo{author}{\bibfnamefont{P.}~\bibnamefont{Thomas}},
  \bibinfo{journal}{Physical Review B} \textbf{\bibinfo{volume}{55}},
  \bibinfo{pages}{7460} (\bibinfo{year}{1997}).

\bibitem[{\citenamefont{Caravaca et~al.}(2009)\citenamefont{Caravaca, Voje,
  Bergli, Ortu\~no, and Somoza}}]{CaVo09}
\bibinfo{author}{\bibfnamefont{M.}~\bibnamefont{Caravaca}},
  \bibinfo{author}{\bibfnamefont{A.}~\bibnamefont{Voje}},
  \bibinfo{author}{\bibfnamefont{J.}~\bibnamefont{Bergli}},
  \bibinfo{author}{\bibfnamefont{M.}~\bibnamefont{Ortu\~no}}, \bibnamefont{and}
  \bibinfo{author}{\bibfnamefont{A.~M.} \bibnamefont{Somoza}},
  \bibinfo{journal}{Annalen der Physik} \textbf{\bibinfo{volume}{18}},
  \bibinfo{pages}{873–876} (\bibinfo{year}{2009}).

\bibitem[{\citenamefont{Cugliandolo et~al.}(1997)\citenamefont{Cugliandolo,
  Kurchan, and Peliti}}]{CuKu97a}
\bibinfo{author}{\bibfnamefont{L.~F.} \bibnamefont{Cugliandolo}},
  \bibinfo{author}{\bibfnamefont{J.}~\bibnamefont{Kurchan}}, \bibnamefont{and}
  \bibinfo{author}{\bibfnamefont{L.}~\bibnamefont{Peliti}},
  \bibinfo{journal}{Physical Review E} \textbf{\bibinfo{volume}{55}},
  \bibinfo{pages}{3898} (\bibinfo{year}{1997}).

\bibitem[{\citenamefont{Kurchan}(2005)}]{Ku05}
\bibinfo{author}{\bibfnamefont{J.}~\bibnamefont{Kurchan}},
  \bibinfo{journal}{Nature} \textbf{\bibinfo{volume}{433}},
  \bibinfo{pages}{222} (\bibinfo{year}{2005}).

\bibitem[{\citenamefont{Grempel}(2004)}]{Gr04a}
\bibinfo{author}{\bibfnamefont{D.~R.} \bibnamefont{Grempel}},
  \bibinfo{journal}{Europhysics Letters} \textbf{\bibinfo{volume}{66}},
  \bibinfo{pages}{854} (\bibinfo{year}{2004}).

\bibitem[{\citenamefont{Kolton et~al.}(2005)\citenamefont{Kolton, Grempel, and
  Dominguez}}]{KoGr05}
\bibinfo{author}{\bibfnamefont{A.~B.} \bibnamefont{Kolton}},
  \bibinfo{author}{\bibfnamefont{D.~R.} \bibnamefont{Grempel}},
  \bibnamefont{and}
  \bibinfo{author}{\bibfnamefont{D.}~\bibnamefont{Dominguez}},
  \bibinfo{journal}{Physical Review B} \textbf{\bibinfo{volume}{71}},
  \bibinfo{pages}{024206} (\bibinfo{year}{2005}).

\bibitem{MaSh92}
\bibinfo{author}{\bibfnamefont{S.} \bibnamefont{Marianer}}
  \bibnamefont{and}
  \bibinfo{author}{\bibfnamefont{B.~I.}~\bibnamefont{Shklovskii}},
  \bibinfo{journal}{Physical Review B} \textbf{\bibinfo{volume}{46}},
  \bibinfo{pages}{13100} (\bibinfo{year}{1992}).

\bibitem[{\citenamefont{Tsigankov and Efros}(2002)}]{TsEf02}
\bibinfo{author}{\bibfnamefont{D.~N.} \bibnamefont{Tsigankov}}
  \bibnamefont{and} \bibinfo{author}{\bibfnamefont{A.~L.} \bibnamefont{Efros}},
  \bibinfo{journal}{Physical Review Letters} \textbf{\bibinfo{volume}{88}},
  \bibinfo{pages}{176602} (\bibinfo{year}{2002}).

\bibitem[{\citenamefont{Ma et~al.}(1991)\citenamefont{Ma, Fletcher, Zaremba,
  D'Iorio, Foxon, and Harris}}]{MaFl91}
\bibinfo{author}{\bibfnamefont{Y.}~\bibnamefont{Ma}},
  \bibinfo{author}{\bibfnamefont{R.}~\bibnamefont{Fletcher}},
  \bibinfo{author}{\bibfnamefont{E.}~\bibnamefont{Zaremba}},
  \bibinfo{author}{\bibfnamefont{M.}~\bibnamefont{D'Iorio}},
  \bibinfo{author}{\bibfnamefont{C.~T.} \bibnamefont{Foxon}}, \bibnamefont{and}
  \bibinfo{author}{\bibfnamefont{J.~J.} \bibnamefont{Harris}},
  \bibinfo{journal}{Physical Review B} \textbf{\bibinfo{volume}{43}},
  \bibinfo{pages}{9033} (\bibinfo{year}{1991}).

\end{thebibliography}
\end{document}